\documentclass[twocolumn,showpacs,prbprintnumbers,prb,fleqn,superscriptaddress]{revtex4}
\usepackage{epsfig}
\usepackage{graphicx}
\usepackage{amsfonts}
\usepackage{amsmath}
\usepackage[dvips]{color}
\usepackage{color}

\newcommand{\ct}{\cite}

\newcommand{\bi}{\bibitem}
\newcommand{\be}{\begin{equation}}
\newcommand{\ee}{\end{equation}}
\newcommand{\ba}{\begin{eqnarray}}
\newcommand{\ea}{\end{eqnarray}}

\newcommand{\de}{\delta}

\begin{document}

\title{The three site interacting spin chain in staggered field: Fidelity vs Loschmidt echo}
\author{Uma Divakaran}
\email{udiva@iitk.ac.in}
\affiliation{Department of Physics, Indian Institute of Technology Kanpur,
Kanpur 208 016, India}

\begin{abstract}
We study the the ground state fidelity and the ground state Loschmidt echo of a three site 
interacting XX chain in presence of  a staggered field which exhibits
special types of quantum phase transitions due to change in the topology of the Fermi surface,
apart from quantum phase transitions from gapped to gapless phases. 
We find that on one hand, the  fidelity is able to detect only the boundaries
separating the gapped from the gapless phase; 
it is completely insensitive to  the phase transition from two Fermi points region to four Fermi points
region lying within this gapless phase. On the other hand, Loschmidt echo shows a dip only at  a special point in the entire phase diagram and hence
fails to detect any quantum phase transition associated with the present model. 
We provide appropriate arguments in support of this anomalous behavior.
\end{abstract}

\maketitle

\section{Introduction}

The desire of building a quantum computer to solve quantum problems efficiently
has lead to an immense recent development in the studies of quantum information theory in many body systems and 
its connection to quantum phase transitions.  Many important  quantum information theoretic measures exhibit interesting scaling behavior close to
a quantum critical point  (QCP) of a quantum many body system. One such measure is the ground state quantum
fidelity, which is an overlap of the ground state wave function at two different values of the
parameters of the quantum  Hamiltonian \cite{zanardi06,zhou08,gu10,dutta10,sacramento11,damski12}. 
The fidelity has attracted the attention of the condensed matter physicists in recent years  because  of its ability to detect a quantum critical point
without an a priori knowledge of  the order parameter of the system, which otherwise 
is the conventional way of probing a quantum phase transition (QPT)\cite{sachdev99,chakrabarti96,dutta10}. 
The 	quantum fidelity shows a dip at a QCP while the fidelity susceptibility (which  defines the rate at which the fidelity changes 
for a finite system in the limit
when the two parameters under consideration are infinitesimally close in the parameter space) 
shows a peak right there and has a scaling form given in terms of
some of the exponents associated with the corresponding QPT \cite{gu10}. 
Similarly, the fidelity has been conjectured to exhibit an interesting scaling relation involving the
quantum critical exponents also in the thermodynamically  large system for a finite separation between the parameters \cite{rams11}.

Although  approaches based on the fidelity and the fidelity susceptibility have  been successful
in detecting various types of quantum phase transition points, for example, ordinary critical points separating two gapped phases through
a gapless point\cite{zanardi06,zhou08} or topological QPTs\cite{yang08}
or QPT in Bose Hubbard model \cite{buonsante07}, but its usefulness in a general scenario is not yet fully settled.
The absence of a peak in the fidelity susceptibility when $\nu d>2$ (where $\nu$ is the correlation length exponent 
associated with the QCP of $d$-dimensional system) has been argued in Refs. \onlinecite{nofidelity,manisha12}.
It has also been shown that in the  marginal case ($\nu d=2$),  a sharp dip in the fidelity is absent at the QCP  using the example of 
Dirac points in two dimensions\cite{patel13}.
On the other hand, the presence of a quasi-periodic lattice introduces extra peaks in fidelity susceptibility 
which can not be detected by studying the energy spectrum \cite{manisha12}.
To show one more contradiction, we present a model where the fidelity detects the onset of a gapless quantum critical region 
but not the transition between two different phases within this gapless quantum critical region.

One of the other important quantities 
which bridges a connection between the quantum information theory and QPT is the ground state Loschmidt echo ($L$) \cite{quan06,cucchietti07,mukherjee12}.
The Loschmidt echo  is the measure of the overlap (at an instant $ t$) of the same initial state, 
the ground state of the initial Hamiltonian of a many body system $H(\lambda)$, but
evolving under the influence of the two Hamiltonians, $H(\lambda)$ and $H(\lambda+\delta)$; 
in this sense, it is the dynamical counterpart of the static
fidelity \ct{zanardi06}.
This $L$ also  shows a dip at the QCP, thus enabling us to detect it.
From the viewpoint of quantum information theory, Loschmidt echo can be used to measure the quantum to classical 
transition (or the transition from a pure to the mixed state)  of a qubit  coupled to an 
environment consisting of the many-body system. In this case, it is the interaction between the qubit and
the many body system that changes the parameter $\lambda$ of the Hamiltonian to $\lambda+\delta$ \ct{quan06} . 
The notion  of the $L$ was actually introduced in connection to the quantum to classical transition in quantum chaos 
\cite{peres95,jalabert01,karkuszewski02,cerruti02,cucchietti03} and now extended to various other systems undergoing a QPT
like Ising model \cite{quan06}, Bose-Einstein condensate model \cite{zheng08} and Dicke model\cite{huang09}.
It has also been studied experimentally using NMR experiments \cite{buchkremer00,zhang09,sanchez09}.

In this paper, we point out the inability of the fidelity or the  $L$ to detect certain
special class of QCPs by taking the example of staggered 
transverse field in a three spin interacting spin-1/2 XX chain \cite{titvinidze03}. 
There are several studies which explore the phase transition in similar or
slightly different three site interacting Hamiltonians using
tools like zero and finite temperature magnetization and magnetization susceptibility \cite{titvinidze03,threespin1}
and transport properties like spin Drude weight and thermal Drude weight \cite{threespin1,lou04}.
The effect of three site interaction has also been studied using magnetocaloric effect \cite{topilko12} or
by using dynamic structure factors of the ground state \cite{krokhmalskii08}. On the other hand, the effect of three site
interacting Hamiltonian is relatively less explored using quantum information theoretic measures which is the
focus of the present paper. There have been studies using Renyi entropy \cite{eloy12}, concurrence, 
geometric discord and quantum discord \cite{cheng12},
average fidelity of state transfer \cite{hu12}, geometric phase \cite{zhang12}, 
$L$ \cite{lian11} and fidelity \cite{liu12} of different versions of three site interacting Hamiltonian.
This study is an important addition to the
literature as to the best of our knowledge, the fidelity and the $L$ of a Hamiltonian which also undergoes a very
unique type of quantum phase transition, namely, from two Fermi points to
four Fermi points, has not been investigated before.We find surprising results due to the
combined effect of staggered field and three site interacting term in the
calculations of fidelity and $L$ which are not a priori obvious. 
It is worth mentioning here that many interesting experimental observations, like
magnetic properties of solid ${}^3\rm{He}$,
have been interpreted as a consequence  of  the presence of multispin interactions \cite{multispin}.

The outline of the paper is as follows: The model and its zero temperature phase diagram 
along with a brief description on the nature of the 
associated quantum phase transitions are presented in section II. 
With an aim to study different QPTs occurring in this model,  we discuss  the
static probe, i.e., the ground state fidelity and fidelity susceptibility in section III, and the dynamic probe given by the ground state 
Loschmidt echo in Section IV.
We summarize our results in the concluding section V.

\section{Model}
In this section, we briefly discuss the ground state phase diagram of 
the one-dimensional three spin interaction Hamiltonian 
in presence of a staggered field $h_s$ given by the Hamiltonian \cite{titvinidze03}

\begin{eqnarray}
H_{TS}&=&- \sum_{j=1}^N  \frac{J}{2}(\sigma_j^x \sigma_{j+1}^x+\sigma_j^y\sigma_{j+1}^y)\nonumber\\
&-&\sum_{j=1}^N \frac{J_3}{4}(\sigma_j^x \sigma_{j+2}^x+\sigma_j^y\sigma_{j+2}^y)\sigma_{j+1}^z\nonumber\\
&-& \sum_{j=1}^N (-1)^j h_s \sigma_j^z,
\end{eqnarray}
where $\sigma$'s are the usual Pauli matrices satisfying the standard commutation relations
and $N$ is the system size.
Performing the  Jordan Wigner Fermionization from spin-1/2 to spinless Fermions $c_j$ \cite{lieb61,bunder99} with the following definitions, 
\begin{eqnarray}
\sigma_j^+=\frac{\sigma_j^x+i\sigma_j^y}2&=&c_j^+\prod_{k=1}^{j-1}(-\sigma_k^z),\nonumber\\
\sigma_j^-=\frac{\sigma_j^x-i\sigma_j^y}2&=&\prod_{k=1}^{j-1}(-\sigma_k^z) ~c_j,\nonumber\\
\sigma_j^z&=&2c_j^{\dagger}c_j-1,\nonumber
\end{eqnarray}
we get
\begin{eqnarray}
H_{TS}&=&-\frac{J}{2}\sum_j(c_j^{\dagger}c_{j+1}+c_{j+1}^{\dagger}c_j) +\frac{J_3}{4} 
\sum_j (c_j^{\dagger}c_{j+2}+c_{j+2}^{\dagger}c_j)\nonumber\\
&-&\sum_j (-1)^jh_s(c_j^{\dagger}c_j-\frac{1}{2}).
\label{eq_jw1}
\end{eqnarray}
\noindent Here $c_{j}$ is the Fermion annihilation operator at the site $j$.
To diagonalize the above Hamiltonian, it is  convenient to introduce two types of 
spinless fermions on the odd and even sub-lattices as shown below:
$$c_{2j-1}=a_{j-1/2}~{\rm and}~ c_{2j}=b_j.$$
Substituting this in Eq. \ref{eq_jw1} and  performing  Fourier transformation  we obtain,
\begin{eqnarray}
H_{TS}&=&\sum_k H_k\nonumber\\
&=&\sum_k [ 
\epsilon_a(k) a_k^{\dagger}a_k + \epsilon_b(k) b_k^{\dagger}b_k
+\epsilon_{ab}(k) (a_k^{\dagger}b_k +b_k^{\dagger}a_k)
]
\nonumber
\end{eqnarray}
where
\begin{eqnarray}
\epsilon_a(k)&=&\frac{J_3}{2} \cos k+h_s,~~\epsilon_b(k)=\frac{J_3}{2} \cos k-h_s\nonumber\\
\epsilon_{ab}(k)&=&-J \cos(k/2).
\end{eqnarray}
and $k=4\pi n/N$ with $-N/4<n<N/4$ for periodic boundary conditions \cite{derzhko09}.

To make the subsequent calculations of the fidelity and the $L$ more transparent,  we introduce
a set  of  basis vectors given by
$|0,0\rangle,~ |a_k,0\rangle,~ |0,b_k\rangle,~ |a_k,b_k\rangle$
where the first index represents the presence or absence of the $a_k-$ particle
and the second index denotes those of the $b_k-$particle.
In these basis, the reduced  Hamiltonian $H_k$ is given by

\begin{eqnarray} H_k= \left[ \begin{array}{cccc} 
0 & 0& 0& 0\\
0& \epsilon_a(k) & \epsilon_{ab}(k) & 0\\
0 & \epsilon_{ab}(k) & \epsilon_b(k) &0 \\
0 & 0 & 0 & \epsilon_a(k) + \epsilon_b(k)
\end{array} \right]. 
\label{eq_matrix}
\end{eqnarray}
We note that  to diagonalize the above 
Hamiltonian, only the basis $|a_k,0\rangle$
and $|0,b_k\rangle$ needs to be rotated since the Hamiltonian in the other two basis is
already diagonal, i.e., there is no mixing along these two directions.
Let us denote the two new directions for the two new quasi-particles
$\alpha_k$ and $\beta_k$ as $|\alpha_k,0\rangle$ and 
$|0,\beta_k\rangle$, which diagonalize the total Hamiltonian giving
four eigen energies $0$, $E_-(k)$, $E_+(k)$ and $E_-(k)+E_+(k)(=\epsilon_a(k)+\epsilon_b(k))$ corresponding
to the four eigenstates  $|0,0\rangle$, $|\alpha_k,0\rangle$, $|0,\beta_k\rangle$ 
and $|\alpha_k,\beta_k\rangle$ (=$|a_k,b_k\rangle$), respectively. 
The two new eigen energies with $J$ set to unity are
\begin{eqnarray}
E_{\pm}(k)=\frac{J_3}2 \cos k \pm \sqrt{h_s^2 + \cos^2(k/2)}
\end{eqnarray}
with the corresponding eigenvectors
\begin{eqnarray}
|\alpha_k,0\rangle&=&\cos\frac{\theta_k}2|a_k,0\rangle-\sin\frac{\theta_k}2|0,b_k\rangle\nonumber\\
|0,\beta_k\rangle&=&\sin\frac{\theta_k}2|a_k,0\rangle+\cos\frac{\theta_k}2|0,b_k\rangle,
\label{eq_rotation}
\end{eqnarray}
where 
\begin{eqnarray}
\tan \theta_k(h_s) = \frac{\cos (k/2)}{h_s}.
\label{eq_theta}
\end{eqnarray}
Hence, the Hamiltonian can now be written in a diagonalized form as
$$H_{TS}=\sum_k E_-(k)\alpha_k^{\dagger}\alpha_k+E_+(k) \beta_k^{\dagger}\beta_k.$$
The ground state of this Hamiltonian corresponds to all the modes 
with negative energies
filled and hence depending upon the parameter values of the Hamiltonian, the 
system has various phases. We briefly discuss these phases below:
\begin{itemize}
\item When $h_s > J_3/2$: $E_+(k)>0$ and $E_-(k)<0$ for all $k$, and hence the 
ground state for each mode is $|\alpha_k,0\rangle$ with total 
ground state energy 
$E_g=\sum_k E_-(k).$
This is the Antiferromagnetic (AF) phase.

\item When $\sqrt{J_3^2/4-1}<|h_s|<J_3/2$: Some of the $k$ modes from the 
$E_+(k)$ branch become negative whereas $E_-(k)$ is negative for all the modes. This phase has two Fermi points
arising due to zeros of the $E_+(k)$ branch. In this region, 
the ground state for a given mode can be $|\alpha_k,0\rangle$
or $|\alpha_k,\beta_k\rangle$ depending upon whether $E_+(k)$ branch is 
empty or filled. The total ground state energy in this phase is given by
$$E_g=\sum_k E_-(k) \Theta(-E_-(k))+E_+(k)\Theta(-E_+(k)),$$
where $\Theta$ is the Heaviside function.
We call this phase as spin liquid I (SLI) phase.

\item When $0<|h_s|<\sqrt{J_3^2/4-1}$: In this limit, $E_-(k)$ also crosses zero
for some modes resulting to four Fermi points, two from each branch.
Hence, there are three possible ground states for a given mode $k$
depending upon the signs of the energies $E_\pm(k)$ given by 
 $|\alpha_k,0\rangle$, $|0,0\rangle$ and $|\alpha_k,\beta_k\rangle$.
Once again the ground state energy is the sum over all the modes
with negative energies for each branch as written above. This phase is called spin liquid II (SLII) phase.
\end{itemize}

\begin{figure}
\begin{tabular}{c}
\includegraphics[height=2.2in,angle=-90]{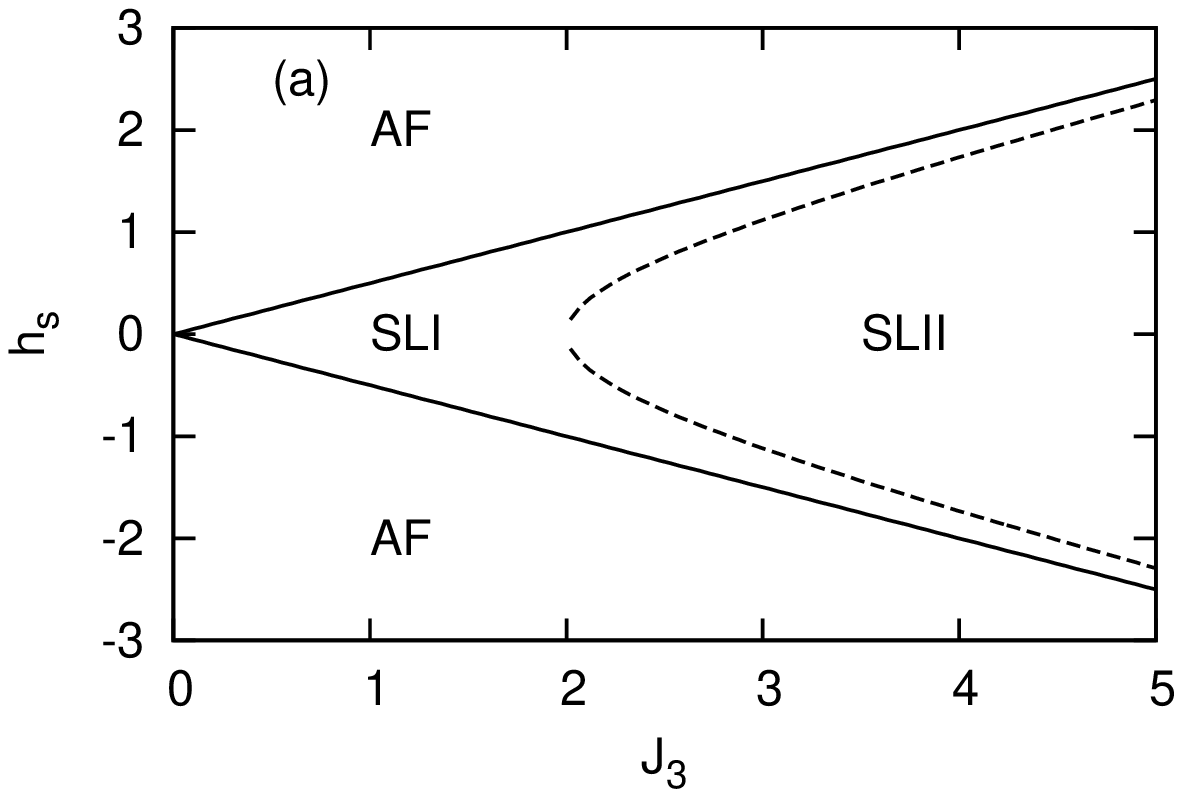}\\
\includegraphics[height=2.2in,angle=-90]{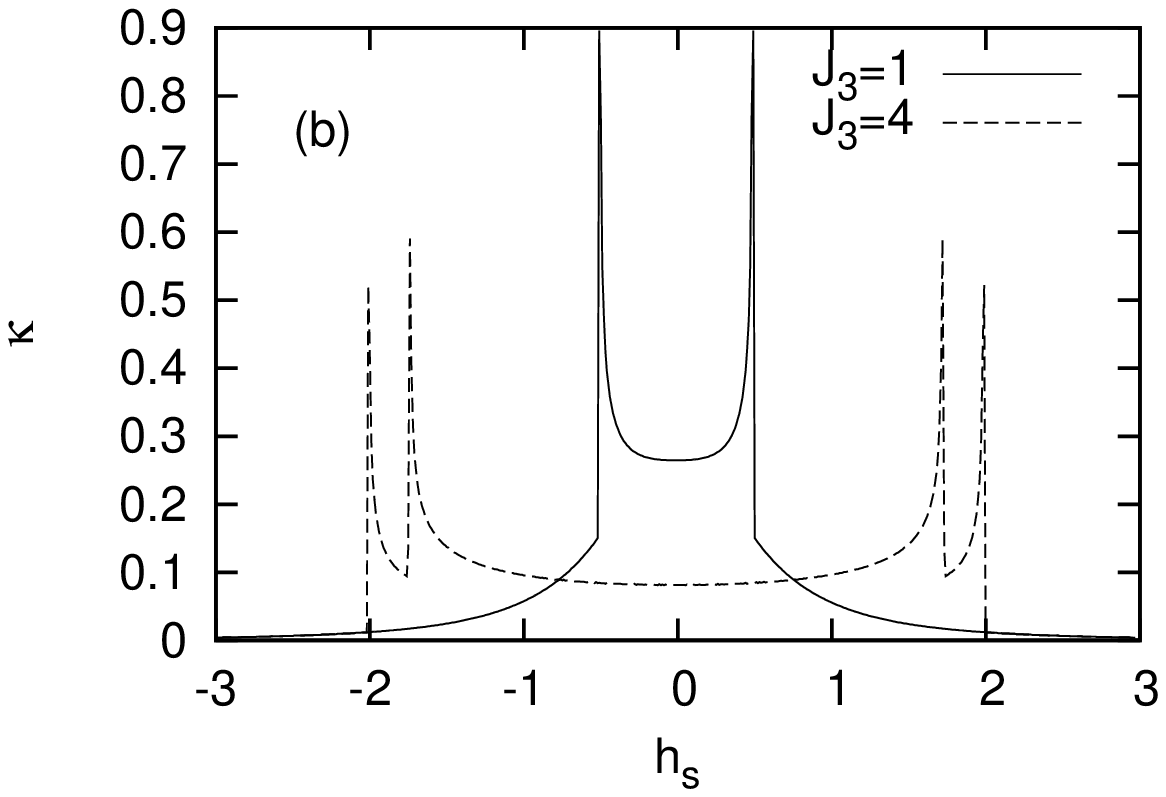}
\end{tabular}
\caption{(a)The phase diagram of the three spin model in presence of a staggered field.
The solid lines correspond to $h_s=\pm J_3/2$ critical line and the dotted lines correspond to
$h_s=\pm \sqrt{(J_3^2-4)/4}$.~(b)The diverging $\kappa$ for two different values of $J_3$. For $J_3=1$,
divergence occurs at $h_s= \pm 0.5$ (AF to SLI) whereas for $J_3=4$, critical points where the divergences occur
are at $h_s=\pm 2$ (AF to SLI) and $h_s=\pm \sqrt{3}$ (SLI to SLII)}
\label{fig_phased}
\end{figure}

Following the above arguments, the phase diagram of the model is shown in 
Fig. \ref{fig_phased} a. 
To complete the discussion on the phase diagram, let us now comment upon the
nature of these quantum phase transitions. We define a stiffness $\kappa$
for a system of size $N$ as
\begin{eqnarray}
\kappa=-\frac{1}{N}\frac{\partial^2 E_g}{\partial h_s^2}.
\end{eqnarray}
A diverging $\kappa$ points to a second order quantum phase transition in the
ground state of the system. In Fig. \ref{fig_phased} b, we present $\kappa$ as a function
of $h_s$ for two different values of $J_3$ which clearly shows that the phase transition
from AF to SLI phase and SLI to SLII phase is indeed a second order quantum phase transition.
The two gapless phases SLI and SLII are characterized by two different types of power law
decays of transverse spin-spin correlation functions 
$C(r)(=\langle \sigma_n^x \sigma_{n+r}^x\rangle=\langle \sigma_n^y \sigma_{n+r}^y\rangle)$ 
as shown in Ref. \onlinecite{titvinidze03}. These correlation functions for $h_s=0$
are given by,
\begin{eqnarray}
C(r)=\frac{A_1}{r^{1/2}}+\frac{B_1\cos(C_1 r)}{r^{5/2}} {\rm {~~~~~in~SLI~ phase}}
\end{eqnarray}
and
\begin{eqnarray}
C(r)&=&\frac{A_2~\cos(C_2 r)}{r}+\frac{B_2 \cos(D_2 r)}{r}+\frac{E_2 \cos(F_2 r)}{r^3}\nonumber\\
&+&\frac{G_2 \cos(H_2 r)}{r^3} {\rm {~~~~~~~~~~~~~~in ~SLII~ phase}},
\end{eqnarray}
where all the constants ($\rm {A_1...H_2}$) are smooth functions of $J_3$.
We now try to capture these phase transitions using the fidelity 
approach, especially  the phase transition from SLI to SLII, and 
later discuss them in the light of Loschmidt echo.

\section{Fidelity}
\begin{figure}[h]
\includegraphics[height=2.1in]{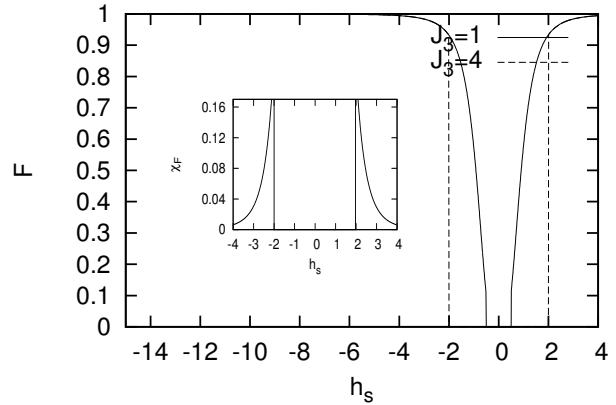}
\caption{The ground state fidelity as a function of the staggered field $h_s$ for two different values of 
$J_3$. The SLII phase exists only for $J_3=4$ but this phase is not captured by the fidelity.
Fidelity only detects the gapped to gapless phase transition of the model which occurs at $h_s=\pm J_3/2$.
Inset shows the fidelity susceptibility for $J_3=4$ as a function of the staggered field for 
the data corresponding to the main figure with a peak at the AF to SLI phase transition
for $h_s=\pm J_3/2$. For better clarity of the figure, we have removed the diverging value of $\chi_F$ at the critical point
as the jump is very large. In both the figures, $N=10^6$ and $\delta=0.01$.}
\label{fig_fidelity}
\end{figure}

As mentioned in the Introduction, the ground state fidelity is defined as the overlap between
the two ground state wave functions at different parameter values; for the present model the fidelity is given by
\begin{eqnarray}
F&=&\langle \Psi_G(h_s)|\Psi_G(h_s+\delta)\rangle
=\prod_k \langle \Phi_k(h_s)|\Phi_k(h_s+\delta)\rangle\nonumber\\
&=&\prod_k F_k,
\label{eq_f}
\end{eqnarray}
where $|\Psi_G(h_s)\rangle$ is the total ground state at $h=h_s$
and $|\Phi_k (h_s)\rangle$ is the ground 
state for the $k-th$ mode. Thus, while 
evaluating $F_k$, one has to carefully  identify the ground state for the
$k-th$ mode depending upon the sign of $E_{\pm}(k)$. The various possibilities
are:
\begin{itemize}
\item $F_k=\langle \alpha_k,0| \alpha_k,0\rangle_{\delta}=\cos(\frac{\theta_k(h_s)-\theta_k(h_s+\delta)}2)$, with $\theta_k$
defined in Eq. \ref{eq_theta}
\item $F_k=\langle \alpha_k,\beta_k |\alpha_k,\beta_k\rangle_{\delta} =\langle 0,0|0,0\rangle_{\delta}=1$
\item  $F_k=\langle \alpha_k,0|\alpha_k,\beta_k\rangle_{\delta}=\langle\alpha_k,0|0,0\rangle_{\delta}=0$
with similar cross products also equal to zero.
\end{itemize}
Here, we fix the notation of the bra/ket without any subscript to denote the  field $h_s$ and with a subscript $\delta$ when  the field is
$h_s+\delta$, which will be followed through out the rest of the paper. We have also used the orthogonality of the basis states
in deriving the above steps.
 It is to be noted 
that within the entire gapless region including SLI and SLII phase, there is 
at least one $k$ mode for which $F_k=0$ and hence fidelity in the entire gapless
region is zero. With these considerations, we numerically evaluate the fidelity using Eq. \ref{eq_f} which is shown in
Fig. \ref{fig_fidelity}. As discussed above, the fidelity is zero in the entire gapless region.
We also briefly comment upon the behavior of the fidelity susceptibility $\chi_F$ which is the second order
derivative of the fidelity with respect to a parameter of the Hamiltonian, and in this case is given as
$$\chi_F =-\frac{\partial^2 F(h_s,\delta)}{\partial h_s^2}.$$
As fidelity, fidelity susceptibility is also not able to capture the SLI to SLII phase transition
and is shown in the inset of Fig. \ref{fig_fidelity}.
To summarize, what we find is that the fidelity (fidelity susceptibility) shows a dip (peak) at the boundary separating
the gapless and gapped phase whereas it fails completely to capture the phase transition occurring inside the gapless phase
which could otherwise be detected by the conventional method of diverging stiffness constant.
We now proceed to study the dynamic counterpart of fidelity, namely Loschmidt echo in the next section.

\section{Loschmidt Echo}


The ground state $L$, defined as the square of the overlap of the initial wavefunction 
given by the ground state of the Hamiltonian $H(h_s)$
but evolving under two different parameter values of the Hamiltonian $h_s$ and $h_s+\delta$, is given as
\begin{eqnarray}
L(t)=|\langle \Psi(h_s,t)| \Psi(h_s+\delta,t)\rangle|^2
\end{eqnarray}
where $|\Psi(h_s,t)\rangle=e^{-iH(h_s)t}|\Psi_G\rangle$ and $|\Psi(h_s+\delta,t)\rangle=e^{-iH(h_s+\delta)t}|\Psi_G\rangle$,
$|\Psi_G\rangle$ being the ground state of the Hamiltonian $H(h_s)$. 
In the momentum representation, as in the case of fidelity,  the expression for the $L$ gets decoupled as
\begin{eqnarray}
L(t)&=& \prod_k |\langle \Phi_k(h_s)|e^{-iH_k(h_s+\delta)t}|\Phi_k(h_s)\rangle|^2\nonumber\\
&=&\prod_k L_k,
\end{eqnarray}
where $|\Phi_k(h_s)\rangle$ is the ground state for the $k-$th mode
at $h=h_s$.
Let us calculate $L_k$ for different possible ground states.
\begin{itemize}
\item If $|\Phi_k(h_s)\rangle=|\alpha_k,0\rangle$
\begin{eqnarray}
L_k= \left|\langle \alpha_k,0 | e^{iH_k(h_s+\delta)t}|\alpha_k,0\rangle\right|^2
\label{eq_le2}
\end{eqnarray}
Since $|\alpha_k,0\rangle$ is the eigenstate of $H_k(h_s)$ and not $H_k(h_s+\delta)$,
we need to rewrite it in terms of $|\alpha_k,0\rangle_{\delta}$, an eigenstate of $H(h_s+\delta)$. 
Using Eq. \ref{eq_rotation},
we find that
\begin{eqnarray}
|\alpha_k,0\rangle=\cos \eta_k |\alpha_k,0\rangle_{\delta} - \sin \eta_k |0,\beta_k\rangle_{\delta},
\end{eqnarray}
where $2\eta_k=\theta_k(h_s)-\theta_k(h_s+\delta)$ and $\theta_k$ is given by Eq. \ref{eq_theta}.
Substituting the above transformation in Eq. \ref{eq_le2}, we get
\begin{eqnarray}
L_k=1-\sin^2(2\eta_k)\sin^2 \left(\frac{\Delta E(k) t}2\right)
\label{eq_LEk}
\end{eqnarray}
with $\Delta E (k)=E_+(h_s+\delta,k)-E_-(h_s+\delta,k)=2\sqrt{(h_s+\delta)^2+\cos^2(k/2)}$.
The three spin term has neither any contribution 
in $\Delta E$ nor in $\eta_k$ and hence does not influence the position of the dip in $L_k$ 
though it will affect the 
magnitude of the dip in $L$ as shown in
Fig. 3; this is because the number of $k-$ modes with $|\alpha_k,0\rangle$ as the ground state
changes as $J_3$ is varied.

\item If $|\Phi_k(h_s)\rangle=|0,0\rangle$ or $|\alpha_k, \beta_k \rangle$, then $L_k=1$ as
these are also the eigen vectors of the Hamiltonian $H_k(h_s+\delta)$.
\end{itemize}
\begin{figure}
\includegraphics[height=3.0in,angle=-90]{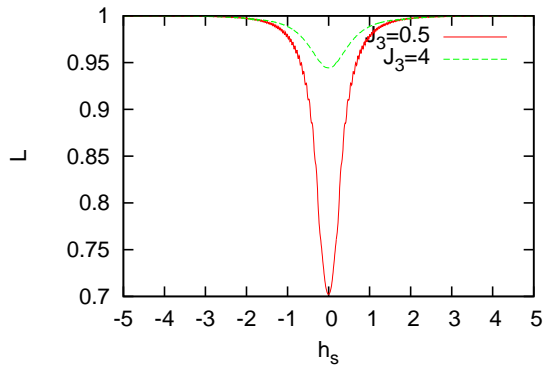}
\caption{Variation of ground state $L$ with the staggered field $h_s$ for two different values of three spin interaction term
with $N=10^6$, $t=100$ and $\delta=0.001$.}
\label{fig_le1}
\end{figure}

From Eq.~(\ref{eq_LEk}), we find that the $L$ is unity deep inside the two antiferromagnetic phases where $2\eta_k$
is infinitesimally small for all $k$ and small $\delta$. The $L$ (or $L_k$) will start deviating from
unity when the term $\sin^2(2\eta_k)\sin^2(\Delta E_k t/2)$ picks up a non-zero value. It can be easily checked
that $2\eta_k$ increases as $\Delta E_k$ approaches zero due to diverging $\tan \theta_k$ (see Eq. \ref{eq_theta}) at $h_s+\delta=0$, 
causing a dip in $L$ only at $h_s+\delta=0$. 
Thus, $L$ can neither detect AF to 
SLI phase transition discussed in section II which is captured by the fidelity, nor it can 
capture SLI to SLII phase transition. Unlike the models studied so far where the modes 
close to the critical mode contribute the most to the decay of the $L$, we will now show that 
it is not so in the present model. 
This is because the ground state for the critical mode
$k=\pi$ at $h_s+\delta=0$ is $|\alpha_k,\beta_k\rangle$, for which $L_{k=\pi}$ is unity.
For the same reasons, the modes close to this critical mode also
do not contribute to the decay of $L$. It is the other modes, away from the critical mode $k=\pi$ at $h_s+\delta=0$ 
satisfying $E_+(k)>0$ and $E_-(k)<0$
which actually influence the behavior of the $L$. This is one of the most interesting observations of the paper.
Since the modes close to the critical mode are not involved in the dynamics, various 
power-law scalings of $L$, which are observed close to the QCP of other models \ct{quan06}, 
are not present in this model.

To understand why $L$ is not able to detect the various ground state phases generated due to
the presence of $J_3$, let us revisit Eq. \ref{eq_matrix}. As mentioned before, $|0,0\rangle$ and $|a_k,b_k\rangle$
do not mix and we can concentrate on $|a_k,0\rangle$ and $|0,b_k\rangle$ basis.
In these two basis, $H_k$ can be written as
\begin{eqnarray}
\frac{J_3}{2} \cos k \hat I + (h_s+\delta) \sigma^z-\cos (k/2) \sigma^x =\frac{J_3}{2} \cos k \hat I + H_{0}
\label{eqn_lz}
\end{eqnarray}
where $\hat I$ is the $2 \times 2$ identity matrix. 
The non-commutativity between the terms involving $\sigma^x$ and $\sigma^z$ causes the time evolution of the spin chain.
On the other hand,
 the identity term in Eq.~(\ref{eqn_lz}) only contributes to the phase of 
the evolving wave function and does not influence the $L$ which is a modulus-squared quantity. 
Hence, the dynamics is dominated by the minimum energy gap of the second part $H_0$ of the Hamiltonian \cite{chowdhury10}, 
which occurs at $h_s+\delta=0$;  consequently, 
the $L$ shows a dip right here. 
Such an observation was also reported
in the context of defect generation for a similar model  where the staggered field is varied linearly as a function of time. 
The study finds  that the scaling of the defect density is insensitive to the QPTs driven by the 
identity operator in the Hamiltonian \cite{chowdhury10}.

\begin{figure}[h]
\begin{tabular}{c}
\includegraphics[height=2.1in,angle=-90]{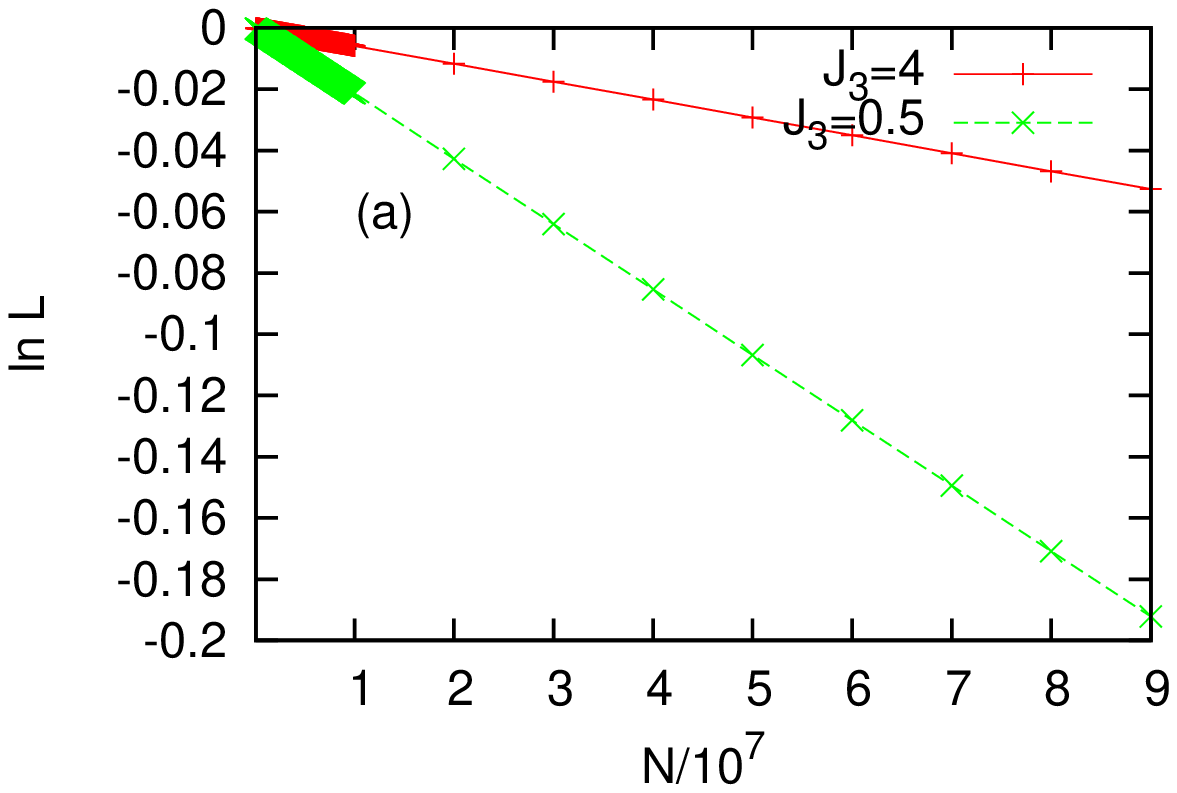}\\
\includegraphics[height=2.1in,angle=-90]{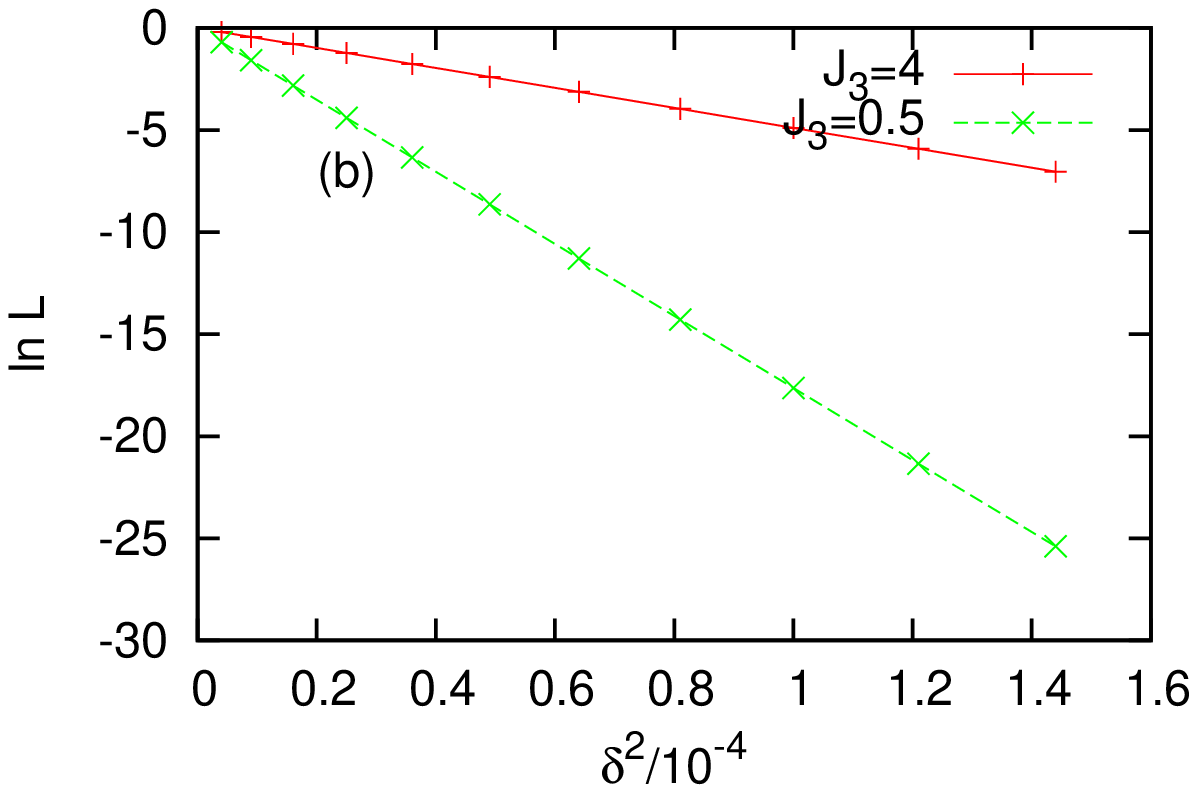}
\end{tabular}
\caption{The exponential decrease of $L$ at $h_s+\delta=0$ with (a) system size $N$ for $t=0.1$ and $\delta=0.001$ and (b) with  
$\delta^2$ for $t=1$ and $N=10^6$.}
\label{fig_le2}
\end{figure}

Since  the $L$  usually shows a  dip in the vicinity of a QCP, one refer to the  staggered field  $h_s+\de=0$ 
as  a  dynamical critical point at which the energy gap vanishes for the dynamical critical mode $k_c^d=\pi$ and 
probe the scaling of the $L$ close to this dynamical critical point.  
We show below that the $L$ decays exponentially with the system size $N$ whereas it decays exponentially with $\delta^2$
for both the regions, $|h_s|>J_3/2$ and $|h_s|<J_3/2$ as shown in Fig. \ref{fig_le2}. 
{To comprehend  this behavior, let us concentrate on
$S=$ln L  which using Eq.~(\ref{eq_LEk}) can be put in the form}
\begin{equation}
S=\sum_k \ln\left(1-\sin^2 2\eta_k \sin^2 \frac{\Delta E_k t}2\right),
\end{equation}
where the summation is only over the relevant $k-$modes satisfying $E_-(k)<0$ and 
$E_+(k)>0$.
Since $\sin^2 2\eta_k$  is
the difference between two approximately equal angles in the limit $\delta \to 0$, it is very small
as the dynamical critical mode and the near by modes do not appear in the summation.
We therefore get a simplified expression
\begin{eqnarray}
S&\simeq& -\sum_k 4 \eta_k^2 \sin^2 \left(\frac{\Delta E_k t}2\right)\nonumber\\
&\simeq&-\sum_k \frac{\delta^2 \sin^2 (t\cos k/2)}{\cos^2 (k/2)}
\label{eq_le1}
\end{eqnarray}
which explains the $\delta^2$ dependence of S or $\ln L$. It is difficult to estimate  the system size dependence 
of the $L$ since the critical mode and the modes close to it do not contribute to the decay and hence no further simplifications can be done.
However,  one can focus  on  the modes closest to the dynamical critical mode $k_c^d=\pi$ for which $E_-(k)<0$ and $E_+(k)>0$
which contribute maximally to $L$. In the early time limit, one gets
$$S\simeq-\sum_k \delta^2 t^2 \propto -N$$
which is consistent with Fig. \ref{fig_le2}. 

Our study shows that the $L$ is not able to detect the various phase transitions in the
ground state phase diagram of the three spin interacting spin chain in presence of a staggered field. These transitions are generated
due to the three spin interacting term $J_3$ which does not influence the dynamics of the Hamiltonian. 
The $L$ does not sense the presence of $J_3$ and detects only the QCPs corresponding to the case $J_3=0$ even when $J_3 \neq 0$.

Although our study is restricted to  a spin-1/2 model, the  excitation spectrum as given in Eq.~(\ref{eqn_lz}) may exist in other models also.
For example, one may consider the Bose-Hubbard model in the hard core limit in the presence of a period two superlattice \cite{hen10} which has a rich
phase diagram containing various phases like 
superfluid, Mott insulator, hole vacuum and particle vacuum phase. We expect  similar results for the fidelity and the $L$  in this hard core boson model also . 

At the end, we mention a couple of works which study the fidelity and $L$  of models 
containing multispin interactions in 
Ising type Hamiltonians \cite{liu12,lian11}.
In ref. \onlinecite{lian11}, the effect of some complicated three spin interaction 
on $L$ is studied and it was  reported that a particular term  of the Hamiltonian does not 
effect the position of the dip
of the  $L$ and modifies  only the sharpness of the decay. 
On the other hand, we have studied here a completely
different model with special types of QCPs and 
presented the appropriate arguments on why
the $L$ is not able to detect the gapped to gapless phase transition points. 
At the same time, the physics of our model is very different from the model studied in 
Ref. \onlinecite{lian11} due to the presence of staggered field
which necessitates the introduction of two types 
of quasiparticles, thus making the problem very different from the
models studied till now in the context of fidelity or Loschmidt echo.

\section{Conclusions}
The ground state fidelity, or equivalently the ground state $L$ shows a dip
at the quantum critical point and thus can in principle be used to determine the phase diagram of any model. 
We show that this is not always the case taking an example of a three spin interacting spin chain in presence of a stagerred field
where the conventional method of divergence of stiffness constant could detect all the critical points
but the method of fidelity and $L$ failed.
While fidelity or fidelity susceptibility can capture
the boundary between the 
gapped to gapless phase transition and is unable to detect the
two Fermi points to four Fermi points phase transition within the gapless region, the Loschmidt echo shows a completely
different picture. It is able to detect only one special point in the entire phase diagram 
and is not able to capture the critical points generated due to the presence of the three spin term. 
This is because the dynamics is entirely governed by the
non-commuting terms of the reduced $2\times 2$ Hamiltonian in $k-$space and the identity matrix containing the 
three spin term only adds a phase to
the wave function evolution. One of the interesting observations of this paper which has never been reported
anywhere else to the best of our knowledge, is that the critical mode and its nearby modes do not contribute anything to the decay of Loschmidt echo.
This is in contrast to the usual scenario where the modes around the critical modes contribute maximum.
We would also like to point out here that in Ref. \onlinecite{nofidelity}, it was shown that the fidelity susceptibility will not be able to detect
any QPT if $\nu d>2$ which is not the case here.

{\bf Acknowledgements:} UD acknowledges fruitful discussions with Shraddha Sharma and Tanay Nag, and thanks Amit Dutta for critically reading the manuscript.

\end{document}